\documentclass{amsart}

\usepackage[a4paper, margin=1in]{geometry}
\usepackage[T1]{fontenc}

\usepackage{amsmath}
\usepackage{amssymb}
\usepackage{amsfonts}
\usepackage{mathtools}
\usepackage{yhmath}
\usepackage{bbm}
\usepackage{dsfont}

\usepackage{graphicx}
\usepackage{xcolor}
\usepackage{tikz}
\usepackage{pgfplots}
\pgfplotsset{compat=1.18}
\usetikzlibrary{positioning, shapes.geometric, patterns}

\usepackage{booktabs}
\usepackage{multirow}
\usepackage{tabularx}
\usepackage{adjustbox}
\usepackage{subcaption}
\usepackage{caption}
\usepackage{multicol}
\usepackage{enumitem}

\usepackage[ruled,vlined]{algorithm2e}
\usepackage{algpseudocode}

\usepackage{hyperref}
\usepackage{url}

\usepackage{amsthm}

\definecolor{darkgreen}{rgb}{0.31, 0.47, 0.26}
\definecolor{lightblue}{rgb}{0.35, 0.65, 0.88}
\renewcommand{\hat}{\widehat}

\makeatletter
\g@addto@macro{\endabstract}{\@setabstract}
\newcommand{\authorfootnotes}{\renewcommand\thefootnote{\@fnsymbol\c@footnote}}
\makeatother

\title[Structure-Aware Stylized Image Synthesis for Robust Medical Image Segmentation]{Structure-Aware Stylized Image Synthesis for Robust Medical Image Segmentation}
\date{}

\begin{document}
\maketitle
\begin{center}
\normalsize
\authorfootnotes
Jie Bao\footnote{1486103897@qq.com}\textsuperscript{1}, Zhixin Zhou\footnote{zhixin0825@gmail.com}\textsuperscript{2},
Wen Jung Li\footnote{wenjli@cityu.edu.hk}\textsuperscript{3},
Rui Luo\footnote{ruiluo@cityu.edu.hk}\textsuperscript{3} 
\par \bigskip

\textsuperscript{1}Huaiyin Institute of Technology \par
\textsuperscript{2}Alpha Benito Research \par
\textsuperscript{3}City University of Hong Kong \par
\par\bigskip
\end{center}

\begin{abstract}
Accurate medical image segmentation is essential for effective diagnosis and treatment planning but is often challenged by domain shifts caused by variations in imaging devices, acquisition conditions, and patient-specific attributes. Traditional domain generalization methods typically require inclusion of parts of the test domain within the training set, which is not always feasible in clinical settings with limited diverse data. Additionally, although diffusion models have demonstrated strong capabilities in image generation and style transfer, they often fail to preserve the critical structural information necessary for precise medical analysis. To address these issues, we propose a novel medical image segmentation method that combines diffusion models and Structure-Preserving Network for structure-aware one-shot image stylization. Our approach effectively mitigates domain shifts by transforming images from various sources into a consistent style while maintaining the location, size, and shape of lesions. This ensures robust and accurate segmentation even when the target domain is absent from the training data. Experimental evaluations on colonoscopy polyp segmentation and skin lesion segmentation datasets show that our method enhances the robustness and accuracy of segmentation models, achieving superior performance metrics compared to baseline models without style transfer. This structure-aware stylization framework offers a practical solution for improving medical image segmentation across diverse domains, facilitating more reliable clinical diagnoses.
\end{abstract}

\section{Introduction}
\label{sec:introduction}
\begin{figure}
	\centering
	\includegraphics[width=1\textwidth]{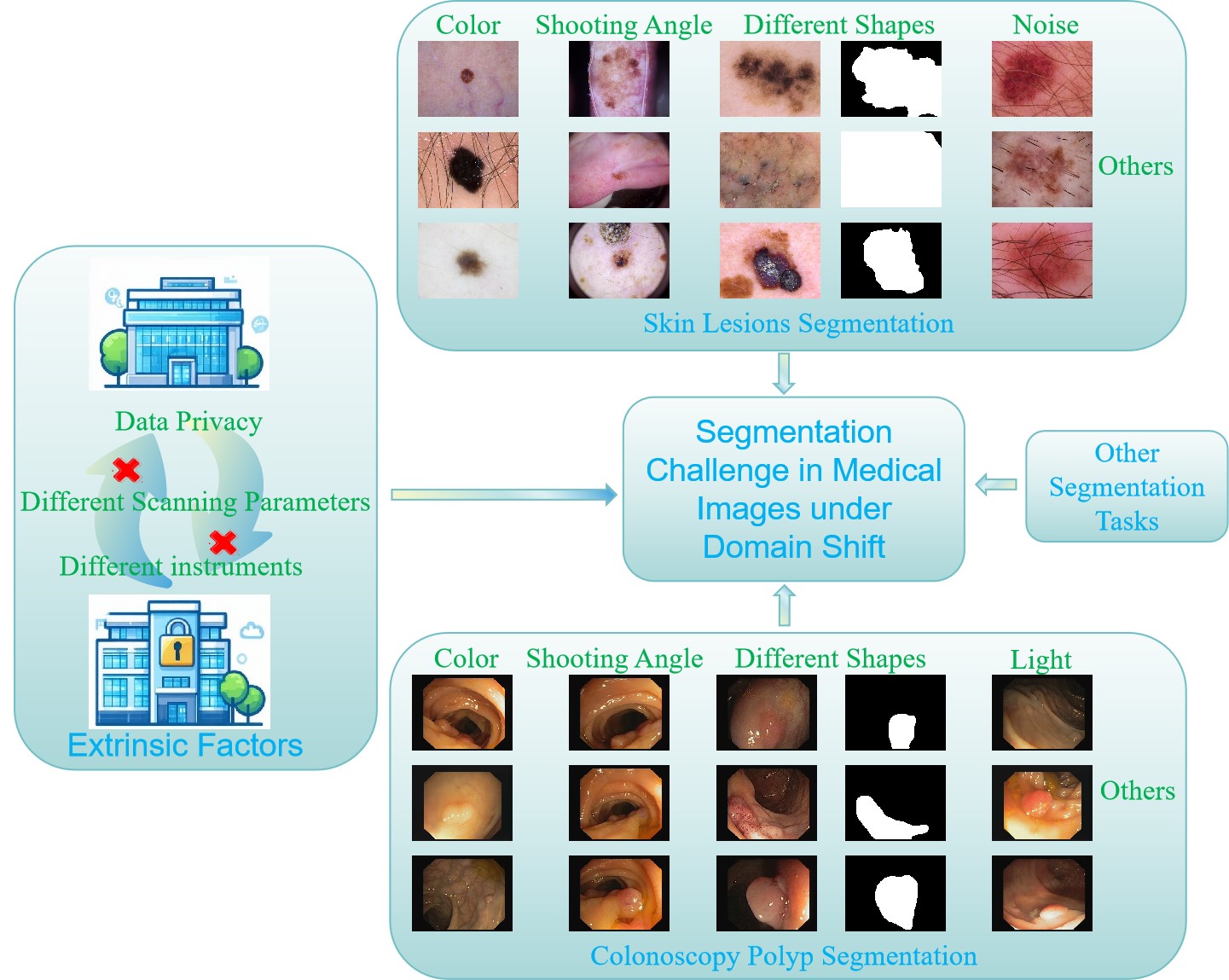}
	\caption{Segmentation Challenge in Medical Images under Domain Shift.}
	\label{fig: Segmentation Challenge}
\end{figure}

Deployment of deep learning models in medical image segmentation faces substantial challenges stemming from variations in imaging devices, acquisition conditions, and patient-specific attributes \cite{lyu2023local, minaee2021image}. As shown in Figure \ref{fig: Segmentation Challenge}, discrepancies arise due to differing imaging modalities, equipment specifications, and environmental factors during image acquisition, all contributing to significant differences in image characteristics \cite{minaee2021image}. Moreover, patient skin color variations impact the visual representation of medical images, thereby influencing the segmentation performance of models \cite{hosny2023deep}. Additionally, the lesions and polyps among diverse patients demonstrate considerable variability in shape, size, and location, compounding the complexity \cite{banik2020polyp}. These variations, exacerbated by the scarcity or inconsistent annotation of labeled data across sources, impede the development of robust machine learning models. To tackle these multifaceted challenges, the implementation of effective image stylization and adaptation techniques is crucial to ensure consistent model performance across various imaging environments and patient demographics.

In the rapidly advancing field of deep learning, the ability of models to reliably perform in various previously unseen environments is crucial. This capability, known as Domain Generalization (DG), focuses on training predictors that maintain robust performance when applied to test distributions different from those encountered during training \cite{wang2023hierarchical, arpit2022ensemble}. In medical image segmentation, DG methods aim to overcome the diversity challenges posed by varying imaging devices, acquisition conditions, and patient-specific attributes, ensuring consistent model performance across diverse clinical settings \cite{ouyang2022causality, hu2022domain}. However, most DG approaches impose stringent requirements, such as necessitating that the training set encompasses a portion of the test domain, which may not be feasible in practical scenarios.

Style transfer emerges as a more effective strategy, where Diffusion Models demonstrate superior generative capabilities by progressively denoising and generating images akin to those in the target domain \cite{zhang2023inversion, lyu2023local}. However, despite their exceptional performance in image generation, Diffusion Models still face challenges in preserving the original structure of input images \cite{ho2020denoising, kwon2022diffusion, zhang2023inversion}. This implies that when applying Diffusion Models for style transfer, critical structural information in medical images may not be adequately retained, thereby limiting their effectiveness in the field of medical image segmentation.

\begin{figure*}
	\centering
	\includegraphics[width=1\textwidth]{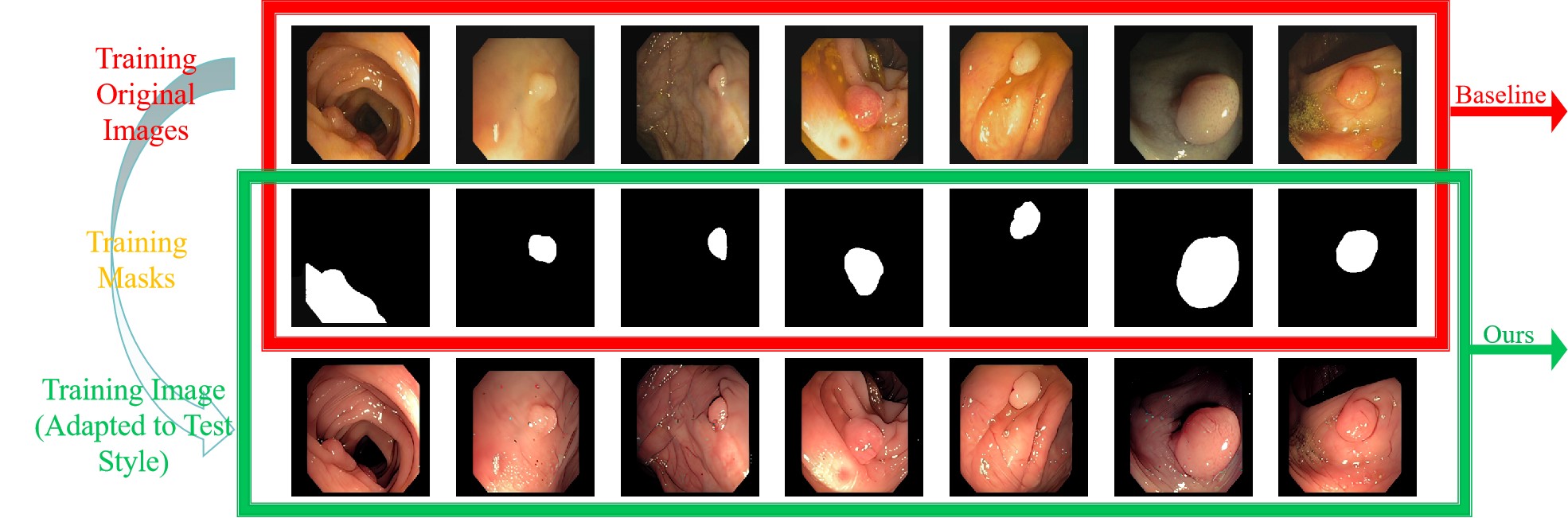}
	\caption{We selected several representative images from the training set in Polyp Datasets, and above are the results after style transfer. As observed, the transferred images still retain the lesion location information corresponding to the mask.}
	\label{fig: style transfer1}
\end{figure*}

To address the aforementioned issues, this paper proposes an image segmentation method based on Diffusion Models for Structure-Aware Single-Shot Image Stylization, aiming to improve the robustness of medical image segmentation models under different devices and acquisition conditions. Our main contributions are summarized as follows:
\begin{itemize}
    \item We propose an effective method to address the domain shift caused by different devices and conditions, which improves the robustness and accuracy of medical image segmentation.
    
    \item As shown in the figure \ref{fig: style transfer1}, our method, incorporating the Structure-Preserving Network (SPN) , ensures the preservation of lesion location and size invariance between the original and stylized images during the style transfer process.
    
    \item The proposed method maintains robust and accurate medical image segmentation performance even when the target domain is not included in the training set.
    
    \item Our approach guarantees that images from other domains in the training set can be successfully transformed into the target domain style.
\end{itemize}

 The manuscript is arranged as follows. Section \ref{Related Works} reviews related work, covering the latest advances in image style transfer and domain generalization; Section~\ref{sec:preliminaries} outlines the problem definition and preliminary knowledge; Section~\ref{Methodology} details our methodology, including the diffusion model, semantic encoder, structure-preserving network (SPN), and segmentation model; Section~\ref{sec:experiments} presents the experiments and results analysis; and Section~\ref{sec:conclusion} concludes the paper with a summary of contributions and future research directions.

\section{Related Works}
\label{Related Works}
\subsection{Image Style Transfer}
Diffusion Models (DMs) have emerged as potent alternatives to traditional generative approaches for image generation and manipulation tasks. While Generative Adversarial Networks (GANs) \cite{goodfellow2014generative} were previously pivotal, they suffered from issues like mode collapse and unstable training. GAN-based methods, such as CycleGAN \cite{zhu2017unpaired}, demonstrated unpaired image-to-image translation but struggled with preserving fine structural details, especially in complex images.

To address these limitations, the research community has increasingly turned to Diffusion Models, which offer a robust framework for high-fidelity image generation by progressively adding noise to data and learning to reverse this process \cite{ho2020denoising}. This iterative denoising procedure inherently captures intricate data distributions, enabling the creation of detailed and realistic images. In the realm of style transfer, diffusion-based approaches have shown significant promise. For instance, DiffuseIT \cite{kwon2022diffusion} leverages the diffusion process to achieve nuanced style transformations while maintaining content consistency. Similarly, InST \cite{zhang2023inversion} employs inversion techniques within the diffusion framework to facilitate precise control over stylization parameters, resulting in high-quality stylized outputs. AStyleDiffusion \cite{wang2023stylediffusion} and ControlNet \cite{zhang2023adding} offer granular control over style transfer, preserving specific structures during complex stylistic changes. Despite this, they sometimes fail to robustly maintain the original image structure, crucial for high-fidelity applications.

One-Shot Structure-Aware Stylized Image Synthesis (OSASIS) \cite{cho2024one} builds upon diffusion models by introducing a Structure Preservation Network (SPN) to maintain spatial integrity during the stylization process. By disentangling semantics from structure, OSASIS ensures that style conversion does not compromise the structural features of the input image, effectively preserving the integrity of the image's original structure. Overall, the integration of diffusion models into style transfer workflows represents a significant advancement, offering enhanced control, stability, and quality over traditional GAN-based approaches.

\subsection{Domain Generalization}
Different imaging devices used by various hospitals can result in discrepancies in the collected images. To reduce dependence on specific domains, adapt models to different scenarios, and enhance their robustness against data distribution shifts, domain generalization (DG) is an efficient and beneficial approach. The goal of domain generalization is to utilize data from multiple source domains during the training phase to learn a model that performs well on unseen target domains. Common DG methods include domain alignment, regularization, meta-learning, and data augmentation.
\begin{enumerate}
    \item \textbf{Domain Alignment} \cite{li2018domain, rame2022fishr, shi2021gradient, sun2016deep}: 
    The core idea of domain alignment is to learn a latent representation space where the distributions of different source domains are as similar as possible. This is achieved by minimizing some measure of distribution discrepancy between domains. For example: \textbf{Domain-Adversarial Neural Network (DANN)} \cite{ganin2016domain} introduces a domain discriminator \(D\) whose goal is to distinguish which source domain the data originates from. The representation function \( \Phi \) and the classifier \( C \) are trained jointly such that \( \Phi \) produces representations that are indistinguishable to \( D \), thereby achieving domain alignment. However, this method may suffer from instability during training. \textbf{Correlation Alignment (CORA)} \cite{sun2016deep} minimizes the difference between covariance matrices of representations from different domains to achieve alignment. \textbf{Domain-Invariant Component Analysis (DICA)} \cite{muandet2013domain} aims to align domains by minimizing differences in higher-order statistics of the latent representations. However, the alignment achieved by these methods may not be very strict.
    \item \textbf{Regularization}: 
    Regularization methods add constraints to the loss function to encourage the model to learn features that are effective across all source domains, thereby achieving domain generalization. Let \( \mathcal{S} = \{\mathcal{S}_1, \mathcal{S}_2, \ldots, \mathcal{S}_N\} \) represent multiple source domains, \( \Phi \) be the representation learning function, and \( C \) be the classifier. Common regularization objectives aim to minimize the risk across all source domains while introducing additional constraints. For example:
    \textbf{Invariant Risk Minimization (IRM)} \cite{arjovsky2019invariant} aims to learn a representation \( \Phi \) such that, for all source domains, the classifier \( w \) is optimal when built on top of \( \Phi \). This method requires multiple optimizations or complex loss calculations, increasing training time. \textbf{Representation Self-Challenging (RSC)} \cite{huang2020self} suppresses dominant features in the training data, encouraging the network to learn other label-correlated features.\textbf{Risk Extrapolation (REx)} \cite{krueger2021out} minimizes the variance between losses across different source domains, hoping that this variance reflects the actual variance in unseen target domains.
    \item \textbf{Meta-Learning}: 
    Meta-learning methods \cite{li2019episodic} emulate the test environment of domain generalization, training models to quickly adapt across different domains. Typically, source domains are split into training and test sets to form "episodes" for multiple training cycles to enhance generalization capabilities. For example: \textbf{Meta-Learning for Domain Generalization (MLDG)} \cite{li2018learning} conducts inner and outer loop optimizations across multiple tasks to learn model parameters that can quickly adapt to new tasks. The model is then evaluated and updated based on performance on a validation set \( S_{\text{val}} \). However, these methods have high training complexity, are sensitive to task design, and require difficult parameter tuning.
    \item \textbf{Data Augmentation}: 
    Data augmentation methods generate more diverse training samples to simulate unseen target domain conditions, thereby improving the model's generalization ability. For example: \textbf{Self-Supervised Learning} \cite{albuquerque2020improving, bucci2021self}: Utilizes different augmentation methods to transform the same image multiple times and matches the losses of these different representations, encouraging the model to learn consistent feature representations. \textbf{Style Transfer} \cite{yue2019domain, zhou2023semi, lyu2023local}: Uses style transfer techniques \cite{wang2023hierarchical, gu2019progressive, zunaed2024learning} to convert source domain images into various styles, simulating the diversity of target domains. In addition, Arpit et al. \cite{arpit2022ensemble} proposed a domain generalization method based on model averaging and ensemble learning. This approach enhances model generalization across different domains by ensembling the moving average parameters from multiple independently trained models. 
\end{enumerate}
However, most domain generalization methods, including domain alignment, regularization, data augmentation, and meta-learning, typically assume that the multiple source domains in the training set can cover or represent the potential test domain distributions. If the test domain significantly differs from the training domains and the training set fails to encompass these differences, the model's generalization ability may substantially deteriorate. 

In the field of medical image segmentation, domain generalization faces numerous challenges:
\begin{itemize}
    \item \textbf{Imaging Equipment Variability}: Different hospitals use various imaging devices \cite{minaee2021image} (e.g., different models of MRI and CT scanners) which can lead to variations in image quality, resolution, and contrast.
    \item \textbf{Patient Demographics}: Variations in patient skin tones can affect the appearance of medical images, influencing the model’s segmentation performance \cite{hosny2023deep}.
    \item \textbf{Lesion Heterogeneity}: Lesions and polyps vary greatly in shape, size, and location across different patients, making it difficult for models trained on one domain to maintain stable performance on another \cite{banik2020polyp}.
    \item \textbf{Annotation Inconsistencies}: Differences in annotation standards and the inherent complexity of cases add to the difficulty of achieving effective domain generalization.
\end{itemize}
Therefore, conducting domain generalization research in medical image segmentation is crucial for developing segmentation models that perform excellently across different hospital equipment and patient populations, thereby improving the accuracy and reliability of clinical diagnoses \cite{ouyang2022causality, hu2022domain}.

To address these challenges, we propose a deep learning-based model that employs the OSASIS method for style transfer and subsequently performs image segmentation to better handle the nonlinear effects between domains. Specifically, the OSASIS method, by incorporating SPN, can perform style transfer on the training set images to target domains while preserving the original spatial structure of medical images, thereby effectively enhancing the model's domain generalization capability. Moreover, the OSASIS model requires only two medical images (one from the source domain and one from the target domain) to achieve the conversion between two domains, and the trained OSASIS model can generalize to perform style transfers from other source domains to the target domain. This characteristic not only addresses the issue of diminished model generalization when there are significant differences between the test and training domains but also allows our style transfer method to be combined with any existing medical image segmentation methods, including UNet \cite{ronneberger2015u}, UNet++ \cite{zhou2018unet++}, PraNet \cite{fan2020pranet}, etc., thereby further improving the stability and accuracy of segmentation performance. Furthermore, the style-transferred images generated by the OSASIS method better preserve the shape and anatomical structure of lesions, facilitating the adaptability of image segmentation models across different hospital devices and patient populations, and enhancing the reliability and accuracy of clinical diagnoses.

\section{Preliminary and Problem Setup}
\label{sec:preliminaries}

The task of medical image segmentation often faces significant challenges due to variations in imaging devices and acquisition conditions. These variations can arise from different imaging modalities (such as MRI versus CT), the types of equipment used, and the settings during image acquisition. When models are applied to new data sources, these differences often degrade their performance, thereby limiting the generalizability and reliability of the models in clinical settings.

To address this issue, we propose the application of style transfer to medical image segmentation. This style transfer method is capable of converting images from different devices and conditions into a consistent style, thereby mitigating the impact of these variations on model performance. Our objective is to transform images from various sources into a unified representation that retains essential structural details while normalizing style-related attributes.

\subsection{Definitions}

\begin{itemize}
    \item \textbf{Source Domain (\( \mathcal{X}_i \)), \( i = 1, 2, \ldots, N \)}: Comprising medical images acquired from \( N \) specific imaging devices or under particular acquisition conditions, where each \( \mathcal{X}_i \) represents a distinct source domain with its own style distribution.
    \item \textbf{Target Domain (\( \mathcal{Y} \))}: Comprising medical images acquired from a different imaging device or under varying acquisition conditions, possessing a unified and distinct style not present in any of the source domains.
    \item \textbf{Mapping Function (\(G\))}: A transformation function \( G: \mathcal{X}_i \rightarrow \mathcal{Y} \) that adapts images from the \( i \)-th source domain to resemble those in the target domain in terms of style and appearance.
    \item \textbf{Structural Integrity Preservation}: Ensuring that the anatomical structures and spatial relationships within the medical images remain minimally altered during the style transfer process.
\end{itemize}

\section{Methodology}
\label{Methodology}
\begin{figure*}
	\centering
	\includegraphics[width=1\textwidth]{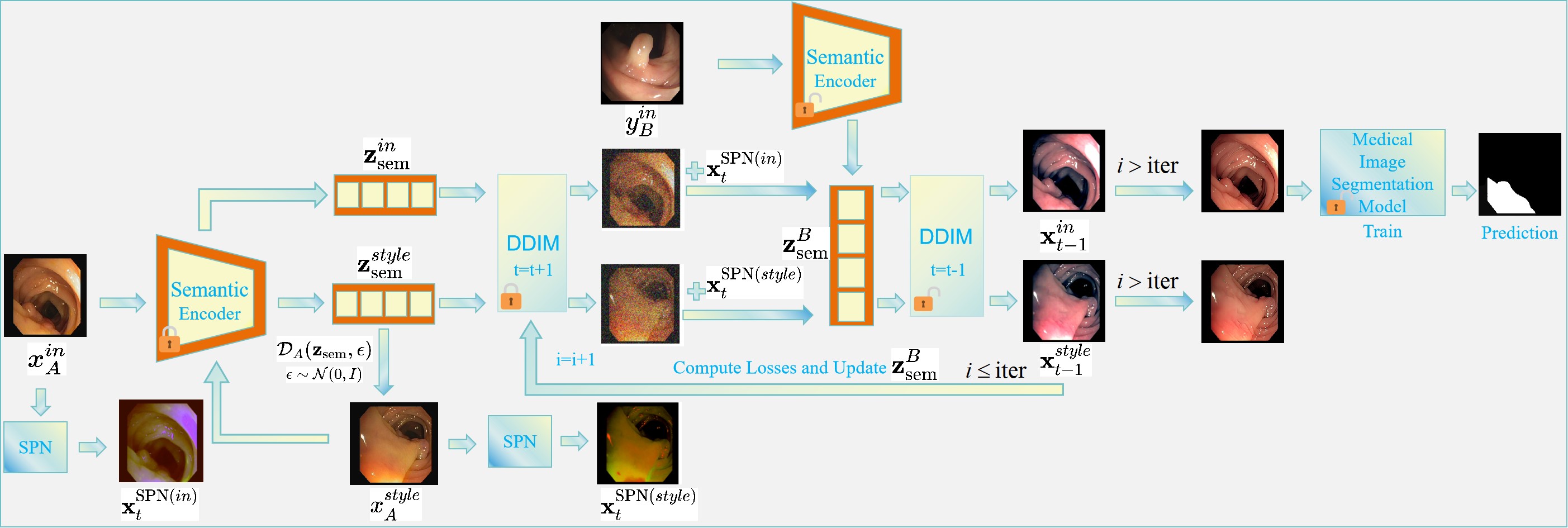}
	\caption{Medical Image Stylization for Segmentation Under Domain Shift.}
	\label{fig: Medical Image Stylization}
\end{figure*}

In this section, we propose an integrated approach for medical image analysis that combines multiple advanced techniques for robust image stylization and accurate segmentation in the context of DG. The methodology is divided into four main components: the Diffusion Model, Semantic Encoder, Structure-Preserving Network (SPN), and the Segmentation Model. Additionally, we define specific loss functions to optimize each component effectively. The overall architecture is illustrated in Figure~\ref{fig: Medical Image Stylization}.
\subsection{Diffusion Model}
\label{subsec:diffusion_model}
Diffusion Models (DMs) \cite{ho2020denoising} have emerged as powerful generative models capable of producing high-fidelity images by modeling the incremental denoising process. Our approach leverages a pretrained Denoising Diffusion Probabilistic Model (DDPM) to facilitate effective image stylization while preserving structural integrity. 

To enhance the efficiency and determinism of the sampling process, we adopt Denoising Diffusion Implicit Models (DDIM) \cite{song2020denoising}. DDIM defines a non-Markovian forward process and derives a corresponding reverse process as follows: 
\begin{gather} 
\label{eq:forward_ddim} 
\mathbf{x}_{t+1} = \sqrt{\alpha_{t+1}} f_\theta(\mathbf{x}_{t}, t) + \sqrt{1 - \alpha_{t+1}} \epsilon_{\theta}(\mathbf{x}_{t}, t), \\ 
\label{eq:reverse_ddim} 
\mathbf{x}_{t-1} = \sqrt{\alpha_{t-1}} f_\theta(\mathbf{x}_{t}, t) + \sqrt{1 - \alpha_{t-1}} \epsilon_{\theta}(\mathbf{x}_{t}, t), 
\end{gather} 
where $f_\theta(\mathbf{x}_t, t)$ is the model's prediction of $\mathbf{x}_{A}^{in}$: 
\begin{gather} 
f_\theta(\mathbf{x}_t, t) = \frac{\mathbf{x}_t - \sqrt{1-\alpha_t} \epsilon_\theta(\mathbf{x}_t, t)}{\sqrt{\alpha_t}}. 
\end{gather} 

\subsection{Semantic Encoding}
\label{subsec:semantic_encoding}

Semantic encoding is crucial for capturing high-level semantic information from input images, which is essential for maintaining content integrity during the stylization process. We utilize the Diffusion Autoencoder (DiffAE) framework \cite{preechakul2022diffusion} to encode input images into semantically rich latent representations.

The Semantic Encoder, denoted as $\operatorname{Enc}_\phi$, encodes a given image $\mathbf{x}_{A}^{in}$ into a semantic latent code $\mathbf{z}_{\mathrm{sem}}$:
\begin{gather}
    \label{eq:semantic_encoder}
    \mathbf{z}_{\mathrm{sem}} = \operatorname{Enc}_\phi(\mathbf{x}_{A}^{in}).
\end{gather}

This latent code $\mathbf{z}_{\mathrm{sem}}$ encapsulates the essential content information of the input image, which is then used to condition the diffusion model during the denoising process. The encoding process aligns the semantic information with the diffusion model, ensuring that the generated images retain the desired content while adapting to the target style. The forward and reverse processes are as follows:
\begin{gather}
    \label{eq:forward_diffae}
    \mathbf{x}_{t+1} = \sqrt{\alpha_{t+1}} f_\theta(\mathbf{x}_{t}, t, \mathbf{z}_{\mathrm{sem}}) + \sqrt{1 - \alpha_{t+1}} \epsilon_{\theta}(\mathbf{x}_{t}, t, \mathbf{z}_{\mathrm{sem}}),
\end{gather}
\begin{gather}
    \label{eq:reverse_diffae}
    \mathbf{x}_{t-1} = \sqrt{\alpha_{t-1}} f_\theta(\mathbf{x}_{t}, t, \mathbf{z}_{\mathrm{sem}}) + \sqrt{1 - \alpha_{t-1}} \epsilon_{\theta}(\mathbf{x}_{t}, t, \mathbf{z}_{\mathrm{sem}}),
\end{gather}
\begin{gather}
    \quad \text{where} \quad f_\theta(\mathbf{x}_t, t, \mathbf{z}_{\mathrm{sem}}) = \frac{\mathbf{x}_t - \sqrt{1-\alpha_t}\epsilon_\theta(\mathbf{x}_t, t, \mathbf{z}_{\mathrm{sem}})}{\sqrt{\alpha_t}} \nonumber.
\end{gather}
   
This process ensures that the semantic information is preserved throughout the diffusion process, enabling effective stylization without compromising the structural integrity of the original image.

\subsection{Structure-Preserving Network (SPN)}
\label{subsec:spn_structure}

Preserving the structural integrity of medical images during stylization is paramount to ensure that critical anatomical details remain intact. To address this, we introduce the Structure-Preserving Network (SPN), which mitigates the loss of structural information introduced during the diffusion process.

The SPN operates by applying a $1 \times 1$ convolution to the intermediate latent representations, effectively preserving spatial information. The incorporation of SPN into the reverse diffusion process is formalized as follows:

\begin{gather}
    \label{eq:output_SPN}
    \mathbf{x}_{t}^{\text{SPN}} = \text{SPN}(\mathbf{x}^{\mathrm{in}}_A), \\
    \label{eq:combine_SPN}
    \mathbf{x}'_{t} = \mathbf{x}_{t} + \mathbf{x}_{t}^{\text{SPN}}, \\
    \label{eq:reverse_diffae_in}
    \mathbf{x}_{t-1} = \sqrt{\alpha_{t-1}} f_\theta(\mathbf{x}'_{t}, t, \mathbf{z}_{\mathrm{sem}}^{\mathrm{in}}) + \sqrt{1 - \alpha_{t-1}} \epsilon^B_{\theta}(\mathbf{x}'_{t}, t, \mathbf{z}_{\mathrm{sem}}^{\mathrm{in}}).
\end{gather}

The SPN ensures that the structural features of the input image $\mathbf{x}^{in}_A$ are preserved by directly adding the SPN-modified latent $\mathbf{x}_{t}^{\text{SPN}}$ to the current timestep latent $\mathbf{x}_{t}$. This adjustment helps maintain the spatial and anatomical integrity of the image throughout the denoising process, resulting in stylized images that retain critical structural details necessary for accurate medical analysis.

\subsection{Segmentation Model}
\label{subsec:segmentation_model}

Accurate segmentation of medical images is essential for clinical diagnosis and treatment planning. Modern segmentation architectures typically include advanced modules for feature extraction and boundary refinement, which help enhance the segmentation accuracy.

In our framework, the stylized images generated by the diffusion model are fed into the segmentation model to produce precise segmentation maps. The integration of the segmentation model ensures that the segmentation process benefits from both the enhanced visual quality of the stylized images and the model's capability to accurately delineate anatomical structures. This combination facilitates accurate delineation of regions of interest, such as lesions or abnormalities, across different imaging domains.

\subsection{Loss Functions}

The training of the OSASIS and PraNet frameworks involves two distinct categories of loss functions: \textbf{Style Transfer Loss} and \textbf{Image Segmentation Loss}. These loss types are optimized separately to prevent conflicts and ensure that each objective is effectively achieved without interference.

\subsubsection{Style Transfer Loss}

The Style Transfer Loss is designed to enforce the adaptation of image styles while preserving structural integrity. It comprises the following components:

\paragraph{Adversarial Loss (\( \mathcal{L}_{\text{adv}} \))}

The adversarial loss ensures that the stylized images align with the target domain's style. Utilizing CLIP directional loss \cite{zhu2021mind}, the framework encourages the semantic latent \( \mathbf{z}_{\text{sem}} \) to capture style-related attributes without compromising structural details:

\begin{align}
    \mathcal{L}_{\text{adv}} = \mathcal{L}_{\text{CLIP}}(\mathbf{z}_{\text{sem}}, \mathbf{z}_{\text{style}}).
\end{align}

\paragraph{Cycle Consistency Loss (\( \mathcal{L}_{\text{cycle}} \))}

To ensure structural integrity, a cycle consistency loss is employed. This loss penalizes discrepancies between the original and reconstructed images after performing round-trip stylization:

\begin{equation}
\begin{split}
    \mathcal{L}_{\text{cycle}} = \mathbb{E}_{\mathbf{x}_{A}^{in} \sim p_{\text{data}}(\mathbf{x}_{A}^{in})} \left[ \| G(F(G(\mathbf{x}_{A}^{in}))) - G(\mathbf{x}_{A}^{in}) \|_1 \right] + \mathbb{E}_{\mathbf{y} \sim p_{\text{data}}(\mathbf{y})} \left[ \| F(G(F(\mathbf{y}))) - F(\mathbf{y}) \|_1 \right],
\end{split}
\end{equation}

where \( F \) and \( G \) denote the forward and reverse stylization processes, respectively.

\paragraph{Structure Preservation Loss (\( \mathcal{L}_{\text{SPN}} \))}

The Structure Preservation Network introduces a loss to ensure that structural details are maintained during stylization:

\begin{align}
    \mathcal{L}_{\text{SPN}} = \mathbb{E}_{\mathbf{x}_{A}^{in} \sim p_{\text{data}}(\mathbf{x}_{A}^{in})} \left[ \| \text{SPN}(\mathbf{x}_{t0}) - \mathbf{x}_{\text{SPN}, t} \|_2 \right].
\end{align}

\paragraph{Total Style Transfer Loss}

The overall style transfer loss is a weighted combination of the above components:

\begin{align}
    \mathcal{L}_{\text{style}} = \lambda_1\mathcal{L}_{\text{adv}} + \lambda_2 \mathcal{L}_{\text{cycle}} + \lambda_3 \mathcal{L}_{\text{SPN}},
\end{align}

where \( \lambda_1 \), \( \lambda_2 \) and \( \lambda_3 \) are hyperparameters balancing the contribution of each loss component.

\subsubsection{Image Segmentation Loss}

The Image Segmentation Loss directly improves the segmentation performance by measuring the discrepancy between the predicted segmentation masks and the ground truth. It is defined as follows:

\paragraph{Segmentation Loss (\( \mathcal{L}_{\text{seg}} \))}

This loss quantifies the difference between the predicted segmentation masks generated by the segmentation model and the corresponding ground truth masks:

\begin{align}
     \mathcal{L}_{\text{seg}} = \mathbb{E}_{\mathbf{X}_{train}, \mathbf{X}_{label}} \left[ \ell(f_\theta(\mathbf{X}_{train}), \mathbf{X}_{label}) \right],
\end{align}

where \( \text{Segment} \) denotes the segmentation model, and \( \mathbf{X}_{label} \) represents the ground truth segmentation masks.

\subsection{Algorithm}

Our approach achieves high-quality style transfer and precise segmentation of medical images in the context of Domain Generalization (DG) by integrating diffusion models, semantic encoders, structure-preserving networks, and segmentation models, combined with carefully designed loss functions. Algorithm \ref{Medical image style transfer} and Algorithm \ref{Segmentation Under Domain Shift} respectively describe the specific implementation processes for style transfer and segmentation, ensuring the effectiveness and robustness of the method.
\begin{algorithm*}
\caption{Medical Image Style Transfer Algorithm Using One-Shot Structure-Aware.}
\label{Medical image style transfer}
\textbf{Input:} {Training Dataset $X_{train}$, Test Dataset $Y_{test}$, Hyperparameters $\lambda_1, \lambda_2, \lambda_3$, Number of Iterations $n$, Forward Diffusion Frequency $T_{1}$, Backward Propagation Time $T_{2}$.} 
\\
\textbf{Output:} {Optimized Mapping Functions $\mathcal{G}$.}
\begin{algorithmic}[1]
\rlap{\hbox{\textcolor{darkgreen}{\Comment{Select Source and Target Images:}}}}
\State Select a random image $x_{A}^{in}\in \mathcal{X}_i$ from $X_{train}$ as Domain $A$ and $y_{B}^{in}\in \mathcal{Y}_i$ from $Y_{test}$ as Domain $B$.

\rlap{\hbox{\textcolor{darkgreen}{\Comment{Initialize Models and Optimizers:}}}}
\State  \parbox[t]{\linewidth}{Load pre-trained CLIP model and DiffAE models for Domains $A$ and $B$;\\
Freeze parameters of DiffAE $\mathcal{D}_A$ for Domain $A$ and Set DiffAE $\mathcal{D}_B$ for Domain $B$ to train mode.}\\
\For{$i = 1$ \textbf{to} $n$}{
    Generate $x_{A}^{style} = \mathcal{D}_A(\mathbf{z}_{\text{sem}}, \mathbf{\epsilon})$ where $\mathbf{z}_{\mathrm{sem}} = \operatorname{Enc}_\phi(x_{A}^{in})$ and $\mathbf{\epsilon} \sim \mathcal{N}(0, I)$ is the random noise;
    
    Encode the style image $\mathbf{z}_{\mathrm{sem}}^{style} = \operatorname{Enc}_\phi(x_{A}^{style})$
    and content image $\mathbf{z}_{\mathrm{sem}}^{in} = \operatorname{Enc}_\phi(x_{A}^{in})$;

    \rlap{\hbox{\textcolor{lightblue}{\Comment{Forward Diffusion Process:}}}}

    \For{$t = 1$ \textbf{to} $T_{1}$}{
    Perform forward diffusion on $x_{\text{t}}^{style}$ and $x_{\text{t}}^{in}$ according to formulas \ref{eq:forward_diffae};

    Based on formulas \ref{eq:output_SPN}, obtain $\mathbf{x}_{t}^{\text{SPN}(style)}$ and $\mathbf{x}_{t}^{\text{SPN}(in)}$ 
    
    for the Structure-Preserving Network.}
    
    \rlap{\hbox{\textcolor{lightblue}{\Comment{Backward Diffusion Process:}}}}
    
    \For{$t = 1$ \textbf{to} $T_{2}$}{Execute Backward Diffusion on $\mathcal{D}_B$:
    
    $\mathbf{x}_{t-1}^{style} = \sqrt{\alpha_{t-1}} f_\theta(x_t^{style} + x_t^{\text{SPN}(style)}, t, \mathbf{z}_{\mathrm{sem}}^B) + \sqrt{1 - \alpha_{t-1}} \epsilon_{\theta}(x_t^{style} + x_t^{\text{SPN}(style)}, t, \mathbf{z}_{\mathrm{sem}}^B)$;

    $\mathbf{x}_{t-1}^{in} = \sqrt{\alpha_{t-1}} f_\theta(x_t^{in} + x_t^{\text{SPN}(in)}, t, \mathbf{z}_{\mathrm{sem}}^B) + \sqrt{1 - \alpha_{t-1}} \epsilon_{\theta}(x_t^{in} + x_t^{\text{SPN}(in)}, t, \mathbf{z}_{\mathrm{sem}}^B)$.}

    \rlap{\hbox{\textcolor{lightblue}{\Comment{Compute Losses:}}}}
    \textbf{Adversarial Loss:}
    \[
    \mathcal{L}_{\text{adv}} =  
    \mathcal{L}_{\text{CLIP}}(\mathbf{z}_{\text{sem}}^A, \mathbf{z}_{\text{sem}}^B);
    \]

    \textbf{Cycle Consistency Loss:} 
    \[
    \mathcal{L}_{\text{cycle}} = \mathbb{E}_{\mathbf{x}_{A}^{in} \sim p_{\text{data}}(\mathbf{x}_{A}^{in})} \left[ \| \mathcal{G}(\mathcal{F}(\mathcal{G}(\mathbf{x}_{A}^{in}))) - \mathcal{G}(\mathbf{x}_{A}^{in}) \|_1 \right] + \mathbb{E}_{\mathbf{y} \sim p_{\text{data}}(\mathbf{y})} \left[ \| \mathcal{F}(\mathcal{G}(\mathcal{F}(\mathbf{y}))) - \mathcal{F}(\mathbf{y}) \|_1 \right];
    \]

    \textbf{Structure Preservation Loss:} 
    \[
    \mathcal{L}_{\text{SPN}} = \mathbb{E}_{\mathbf{x}_{A}^{in} \sim p_{\text{data}}(\mathbf{x}_{A}^{in})} \left[ \| \text{SPN}(\mathbf{x}_{A}^{in}) - \mathbf{x}_{\text{SPN}, t} \|_2 \right];
    \]

    \textbf{Total Style Transfer Loss:} 
    \[
    \mathcal{L}_{\text{style}} =\lambda_1 \mathcal{L}_{\text{adv}} + \lambda_2 \mathcal{L}_{\text{cycle}} + \lambda_3 \mathcal{L}_{\text{SPN}};
    \]
    
    \rlap{\hbox{\textcolor{lightblue}{\Comment{Backpropagation and Optimization:}}}}
    Backpropagate the total loss $\mathcal{L}_{\text{style}}$ and update the Structure-Preserving Network and mapping functions $\mathcal{G}$, specifically $\mathbf{z}_{\mathrm{sem}}^B$.
}
\end{algorithmic}
\end{algorithm*}

\begin{algorithm*}
\caption{Medical Image Stylization for Segmentation Under Domain Shift.}
\label{Segmentation Under Domain Shift}
\textbf{Input:} {Training Dataset $X_{train}$, Training Dataset Label $X_{label}$, Test Dataset $Y_{test}$, Mapping Functions $\mathcal{G}$.} 
\\
\textbf{Output:} {Predicted segmentation masks $\{\hat{y}_j\}$.}
\begin{algorithmic}[1]
\State Initialize model parameters $\theta$ and Image Segmentation Model $f_{\theta}$. \\ 
\For{each batch $x_i$ in $X_{train}$}{
        $x'_i = \mathcal{G}(x_i)$. \textcolor{darkgreen}{\Comment{\footnotesize Apply style transfer.}}} \\  
\For{epoch = 1 to N}
    { $\hat{y}_i = f_{\theta}(x'_i)$ ; 
    
$\mathcal{L}_{\text{seg}} = \mathbb{E}_{\mathbf{X}_{train}, \mathbf{X}_{label}} \left[ \ell(f_\theta(\mathbf{X}_{train}), \mathbf{X}_{label}) \right]$; \textcolor{darkgreen}{\Comment{\footnotesize Segmentation loss.}}

 Update Image Segmentation Model $f_{\theta}$.}\\
\For{each image $y_j$ in $Y_{test}$}
    {$\hat{y}_j = f_{\theta}(y_j)$.\textcolor{darkgreen}{\Comment{\footnotesize Predict segmentation masks.}}}
\end{algorithmic}
\end{algorithm*}

\section{Experiments}
\label{sec:experiments}
In this section, we evaluate the proposed OSASIS framework on multiple medical image segmentation datasets that exhibit significant variations due to different imaging devices and acquisition conditions. We compare the performance of three base models: U-Net \cite{ronneberger2015u}, U-Net++ \cite{zhou2018unet++}, and PraNet \cite{fan2020pranet}, to demonstrate the effectiveness of our approach in maintaining structural fidelity and enhancing segmentation accuracy across diverse imaging conditions.

\subsection{Baseline Models}
We evaluate three baseline segmentation models: 
\begin{enumerate} 
\item \textbf{U-Net} \cite{ronneberger2015u}: A seminal convolutional network architecture for biomedical image segmentation, known for its symmetric encoder-decoder structure with skip connections. 
\item \textbf{U-Net++} \cite{zhou2018unet++}: An enhanced version of U-Net that integrates dense skip connections and nested U-Nets to capture better feature representations. 
\item \textbf{PraNet} \cite{fan2020pranet}: A U-Net based polyp recognition and attention network that leverages reverse attention and parallel partial decoders for medical image segmentation. 
\end{enumerate} 
For each baseline model, we evaluate the efficacy of our style transfer method by comparing the following two experimental approaches: 
\begin{enumerate} 
\item \textbf{Direct approach}: Training the segmentation model and testing it directly on the test set. 
\item \textbf{Style transfer approach}: First performing style transfer from the training set to match the test set style, then training the segmentation model for testing. 
\end{enumerate}

\subsection{Training Settings}
All models are implemented in PyTorch and trained on an NVIDIA TITAN RTX GPU with 16GB memory.\footnote{The experimental data and content for this work can be found at \url{https://github.com/luo-lorry/Stylized-Medical-Segmentation}.} The input images are uniformly resized to $256 \times 256$ pixels. All models are trained with a batch size of 12. The optimization is performed using the Adam optimizer.
\subsection{Evaluation Metrics}

We employ a comprehensive set of evaluation metrics to assess the performance of the segmentation models: 

\begin{itemize} 
    \item \textbf{Mean Dice Coefficient} and \textbf{Mean Intersection over Union (IoU) \cite{jha2019resunet++}}: Two fundamental metrics that measure the overlap between predicted segmentation and ground truth.
    
    The Dice Coefficient is defined as:
    \[
    \text{Dice} = \frac{2 |A \cap B|}{|A| + |B|},
    \]
    where \( A \) is the set of predicted positive pixels and \( B \) is the set of ground truth positive pixels.
    
    The Intersection over Union (IoU) is defined as:
    \[
    \text{IoU} = \frac{|A \cap B|}{|A \cup B|}.
    \]
    
    \item \textbf{Specificity}: Measures the model's ability to correctly identify negative cases, evaluating the accuracy of non-polyp region detection.
    
    Specificity is calculated as:
    \[
    \text{Specificity} = \frac{TN}{TN + FP},
    \]
    where \( TN \) is the number of true negatives and \( FP \) is the number of false positives.
    
    \item \textbf{Weighted F-measure ($F_\beta^w$)}: An enhanced version of the Dice metric that addresses the "Equal-importance flaw" by assigning different weights to different regions.
    
    The Weighted F-measure is given by:
    \[
    F_\beta^w = (1 + \beta^2) \cdot \frac{\mathrm{Precision} \cdot \mathrm{Recall}}{(\beta^2 \cdot \mathrm{Precision}) + \mathrm{Recall}},
    \]
    where weights are applied to balance precision and recall according to the parameter \( \beta \).
    
    \item \textbf{Structure Measure ($S_\alpha$) \cite{fan2017structure}}: Evaluates the structural similarity between predictions and ground truths, complementing pixel-wise evaluation metrics.
    
    The Structure Measure is defined as:
    \[
    S_\alpha = \alpha \cdot S_o + (1 - \alpha) \cdot S_r,
    \]
    where \( S_o \) assesses object-level similarity and \( S_r \) assesses region-level similarity, with \( \alpha \) balancing the two components.
    
    \item \textbf{Enhanced-alignment Measure ($E_\phi^{\text{max}}$) \cite{fan2018enhanced}}: Assesses both pixel-level and global-level similarities between predicted and ground truth segmentations.
    
    The Enhanced-alignment Measure is expressed as:
    \[
    E_\phi = \max_{\phi \in \Phi} \left[ \phi \cdot S_{\phi} + (1 - \phi) \cdot E_{\phi} \right],
    \]
    where \( \phi \) represents the alignment parameter, and \( S_{\phi} \) and \( E_{\phi} \) denote the pixel-level and global-level similarity measures, respectively.
    
    \item \textbf{Mean Absolute Error (MAE)}: Quantifies the pixel-level accuracy by measuring the average magnitude of prediction errors.
    
    The Mean Absolute Error is calculated as:
    \[
    \text{MAE} = \frac{1}{N} \sum_{i=1}^{N} |P_i - G_i|,
    \]
    where \( P_i \) and \( G_i \) are the predicted and ground truth values for the \( i \)-th pixel, respectively, and \( N \) is the total number of pixels.
\end{itemize}

These metrics collectively constitute a comprehensive evaluation framework, encompassing diverse aspects of segmentation performance such as pixel-level precision, structural resemblance, and region-based assessments.

\subsection{Colonoscopy Polyp Segmentation Experiment} 
Colonoscopy is a critical tool for detecting colorectal polyps, but segmenting polyp images presents two major challenges. First, the diversity of polyp characteristics, including variations in size, color, shape, and texture, makes accurate identification difficult. Second, differences in hospital imaging conditions, such as variations in equipment, lighting, and shooting angles, result in inconsistencies in image quality and clarity, further complicating segmentation. These challenges of variability and inconsistency can be mitigated by applying image segmentation following style transfer on the images.

This experiment utilized two datasets: CVC-ClinicDB \cite{bernal2015wm} (training set) and CVC-ColonDB \cite{tajbakhsh2015automated} (test set). We selected twenty images from CVC-ColonDB containing polyps of different types but with consistent style. From the CVC-ClinicDB dataset, we chose 178 images with a different style from the test set. 

\begin{figure*}
	\centering
	\includegraphics[width=1\textwidth]{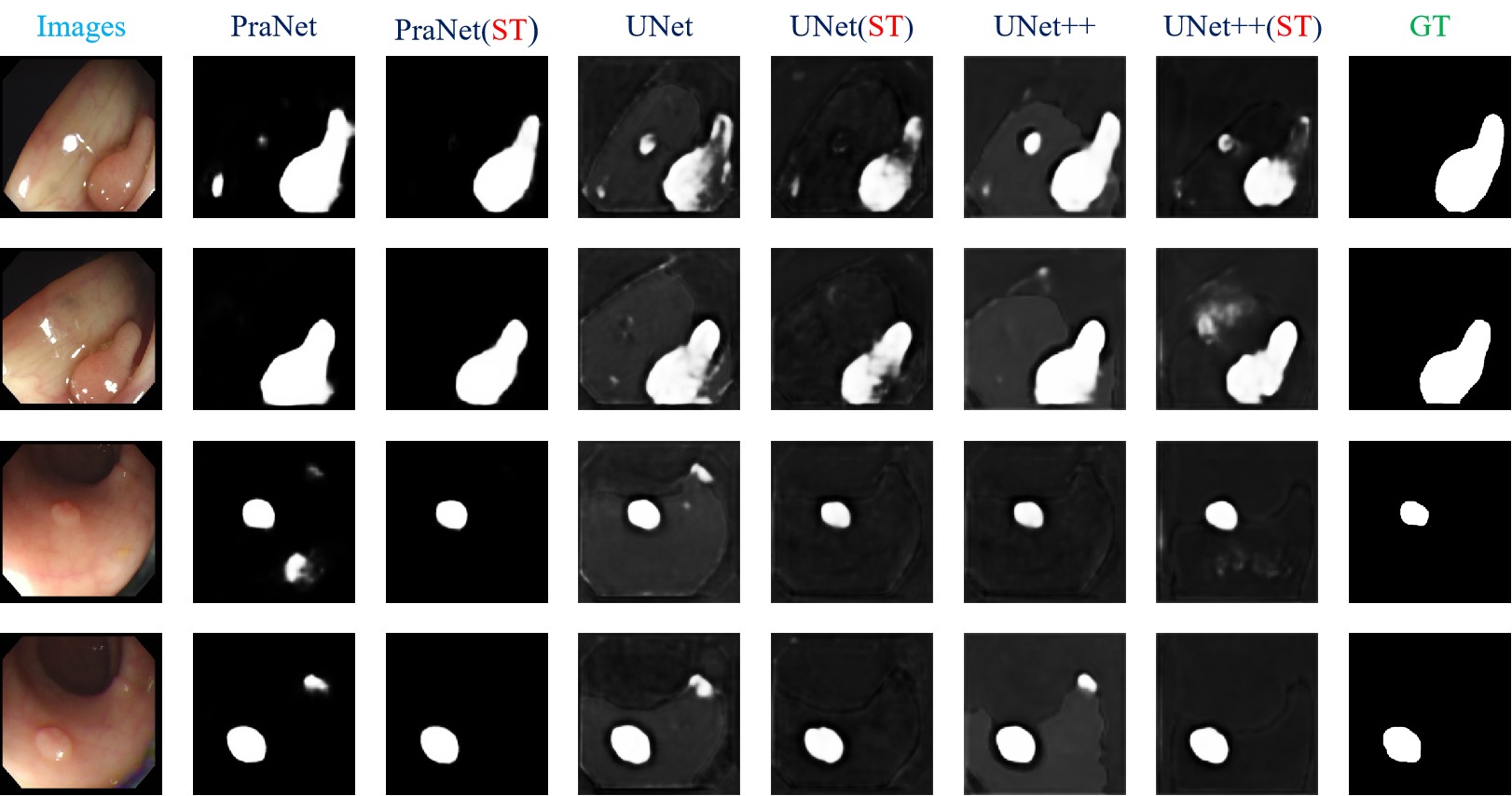}
	\caption{Qualitative results of different models and models after style transfer in Polyp Datasets.}
	\label{fig: results1}
\end{figure*}

\begin{figure*}[htbp]
    \centering
    \begin{minipage}{0.28\linewidth}
        \centering
        \includegraphics[width=\linewidth]{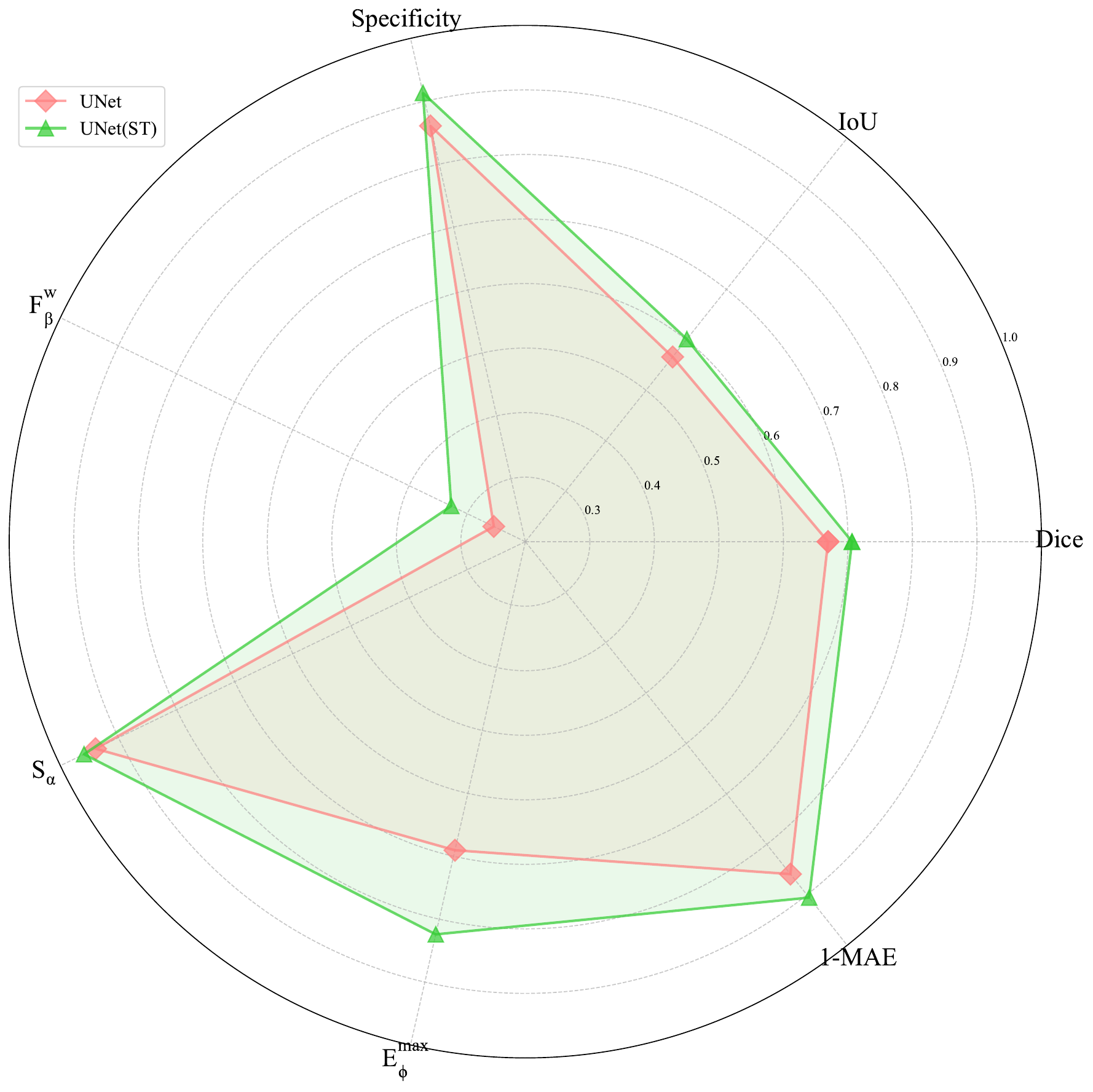}
        \label{fig:figure_Radar1}
    \end{minipage}\hfill
    \begin{minipage}{0.28\linewidth}
        \centering
        \includegraphics[width=\linewidth]{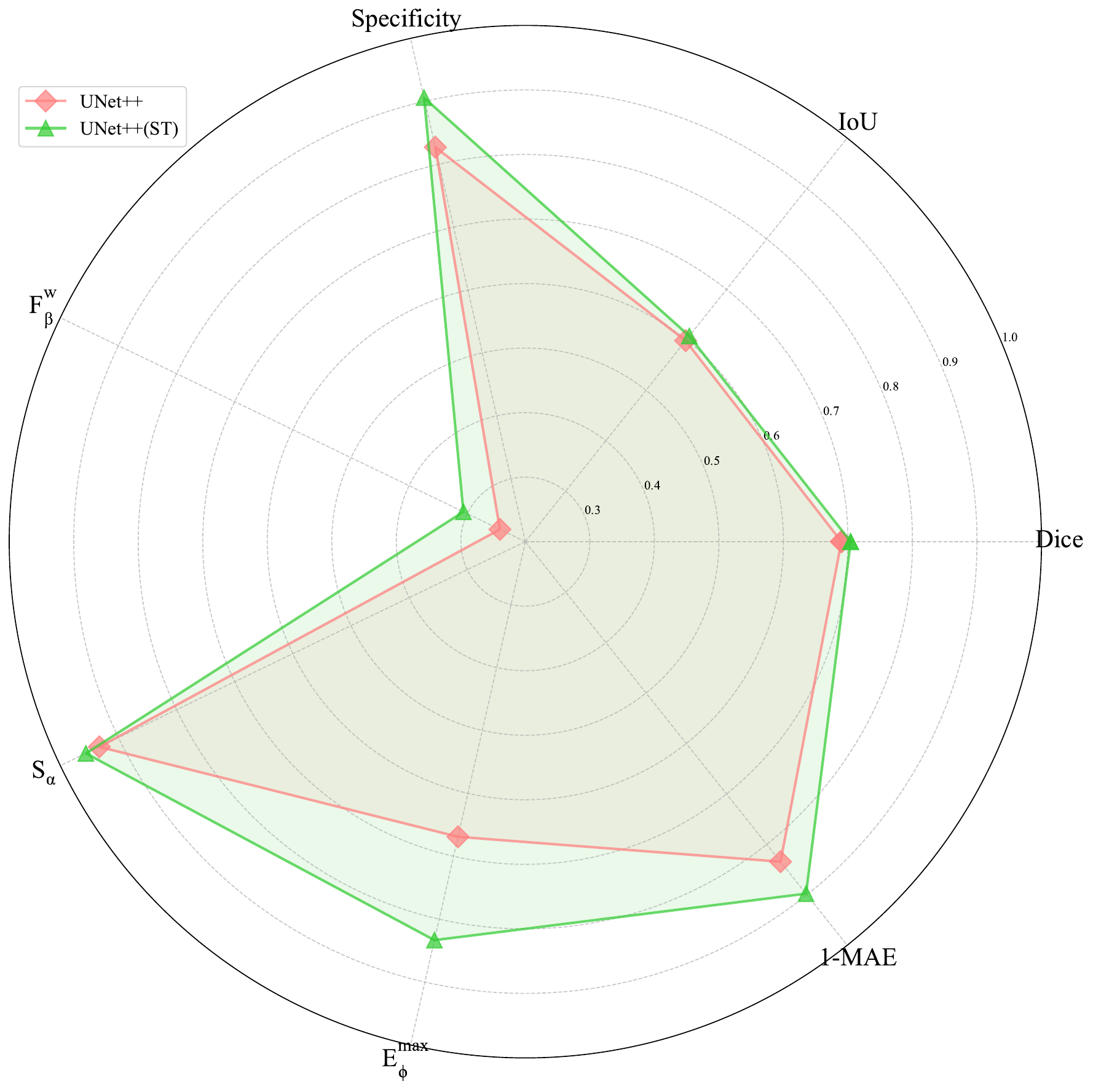}
        \label{fig:figure_Radar2}
    \end{minipage}\hfill
    \begin{minipage}{0.28\linewidth}
        \centering
        \includegraphics[width=\linewidth]{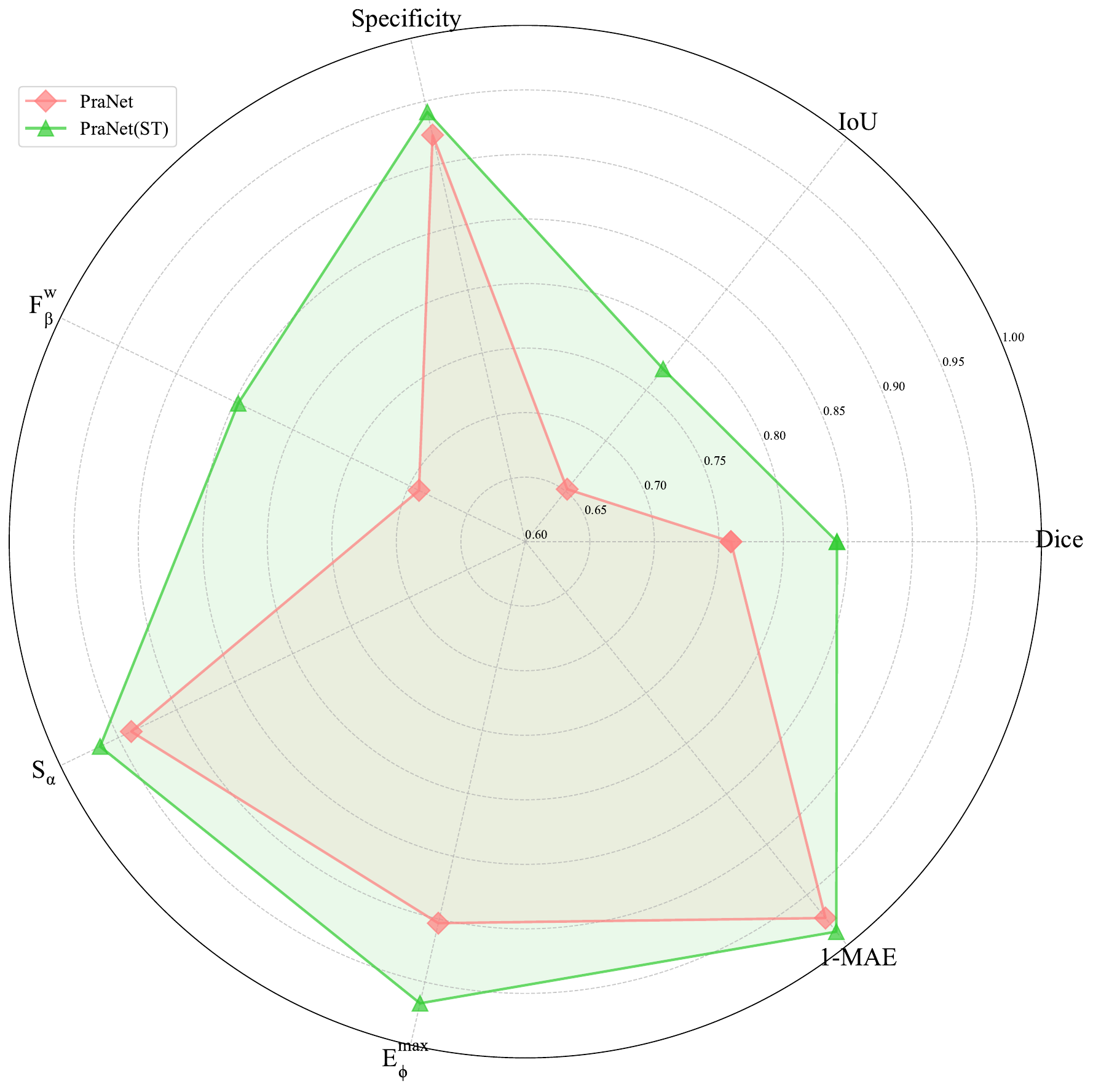}
        \label{fig:figure_Radar3}
    \end{minipage}
    \caption{Radar charts illustrating the performance metrics (Dice, IoU, Specificity, $F_\beta^w$, $S_\alpha$, $E_\phi^{max}$, and 1-MAE) of UNet, UNet++ and PraNet segmentation models and their style transfer variants on Polyp datasets.}
    \label{fig:figure_Radar_Polyp}
\end{figure*}

As shown in Figure \ref{fig: style transfer1}, after applying style transfer to the training set, the images still retain their lesion characteristics, while their styles, including the backgrounds, have been unified. From Table \ref{tab:performance1}, we can observe that compared to the original image segmentation model,  with style transfer before segmentation results in improvements in various metrics on the test set. Specifically, in the PraNet image segmentation model, the Dice coefficient, IoU (Intersection over Union), and $F_\beta^w$ have all increased by approximately 10\%. Furthermore, as illustrated in Figure \ref{fig: results1}, a qualitative analysis demonstrates that applying style transfer to the test set can yield a better image segmentation model, with less noise in the results. Furthermore, we have additionally created radar charts \ref{fig:figure_Radar_Polyp} to more effectively highlight the performance disparities among the various models.

\begin{table}[htbp]
\centering
\adjustbox{width=\linewidth}{%
\begin{tabular}{|c|c|c|c|c|c|c|c|}
\toprule
Model & Dice & IoU & Specificity & $F_\beta^w$ & $S_\alpha$ & $E_\phi^{max}$ & MAE\\
\midrule
UNet & 0.6692 & 0.5661 & 0.8609 & 0.2543 & 0.9394 & 0.6905 & 0.1411\\
UNet(\textbf{ST}) & \textbf{0.7063} & \textbf{0.6014} & \textbf{0.9133} & \textbf{0.3275} & \textbf{0.9592} & \textbf{0.8242} & \textbf{0.0945} \\
\cmidrule{1-8}
UNet++ & 0.6898 & 0.5984 & 0.8270 & 0.2438 & 0.9329 & 0.6691 & 0.1657 \\
UNet++(\textbf{ST}) & \textbf{0.7042} & \textbf{0.6076} & \textbf{0.9057} & \textbf{0.3066} & \textbf{0.9561} & \textbf{0.8334} & \textbf{0.1023}\\
\cmidrule{1-8}
PraNet & 0.7593 & 0.6520 & 0.9232 & 0.6915 & 0.9390 & 0.9032 & 0.0270 \\
PraNet(\textbf{ST}) & \textbf{0.8415} & \textbf{0.7712} & \textbf{0.9415} & \textbf{0.8470} & \textbf{0.9658} & \textbf{0.9669} & \textbf{0.0135} \\
\bottomrule
\end{tabular}}
\caption{Comparison results of different models and their style-transferred versions in terms of performance indicators.}
\label{tab:performance1}
\end{table}

\subsection{Skin Lesions Segmentation Experiment}
Skin lesion segmentation is crucial for early diagnosis of skin diseases, assisting doctors in formulating effective treatment plans, and monitoring disease progression, making it vital in dermatology. However, this task faces two major challenges. First, the diversity in skin tones, variability in lesion colors, and interference from body hair increase the difficulty of accurate segmentation. Second, differences in hospital imaging conditions, such as variations in equipment, lighting, and shooting angles, result in inconsistencies in image quality and clarity, further complicating the segmentation process.

In this experiment, we selected 200 images of skin lesions with pink or red backgrounds from the HAM10000 dataset \cite{tschandl2018ham10000}. Additionally, we chose 20 images of skin lesions with white backgrounds.

\begin{figure*}
	\centering
	\includegraphics[width=1\textwidth]{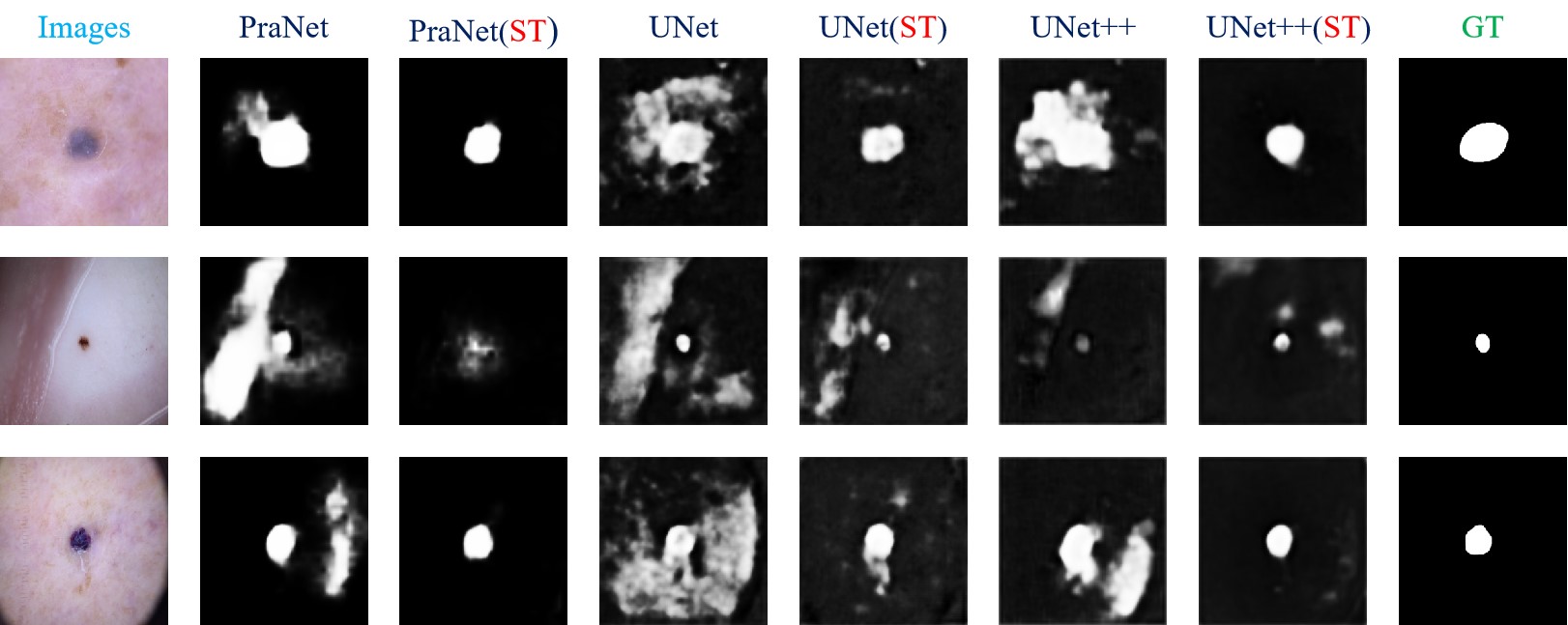}
	\caption{Qualitative results of different models and models after style transfer in Skin Lesions Datasets.}
	\label{fig: results2}
\end{figure*}

\begin{figure*}[htbp]
    \centering
    \begin{minipage}{0.28\linewidth}
        \centering
        \includegraphics[width=\linewidth]{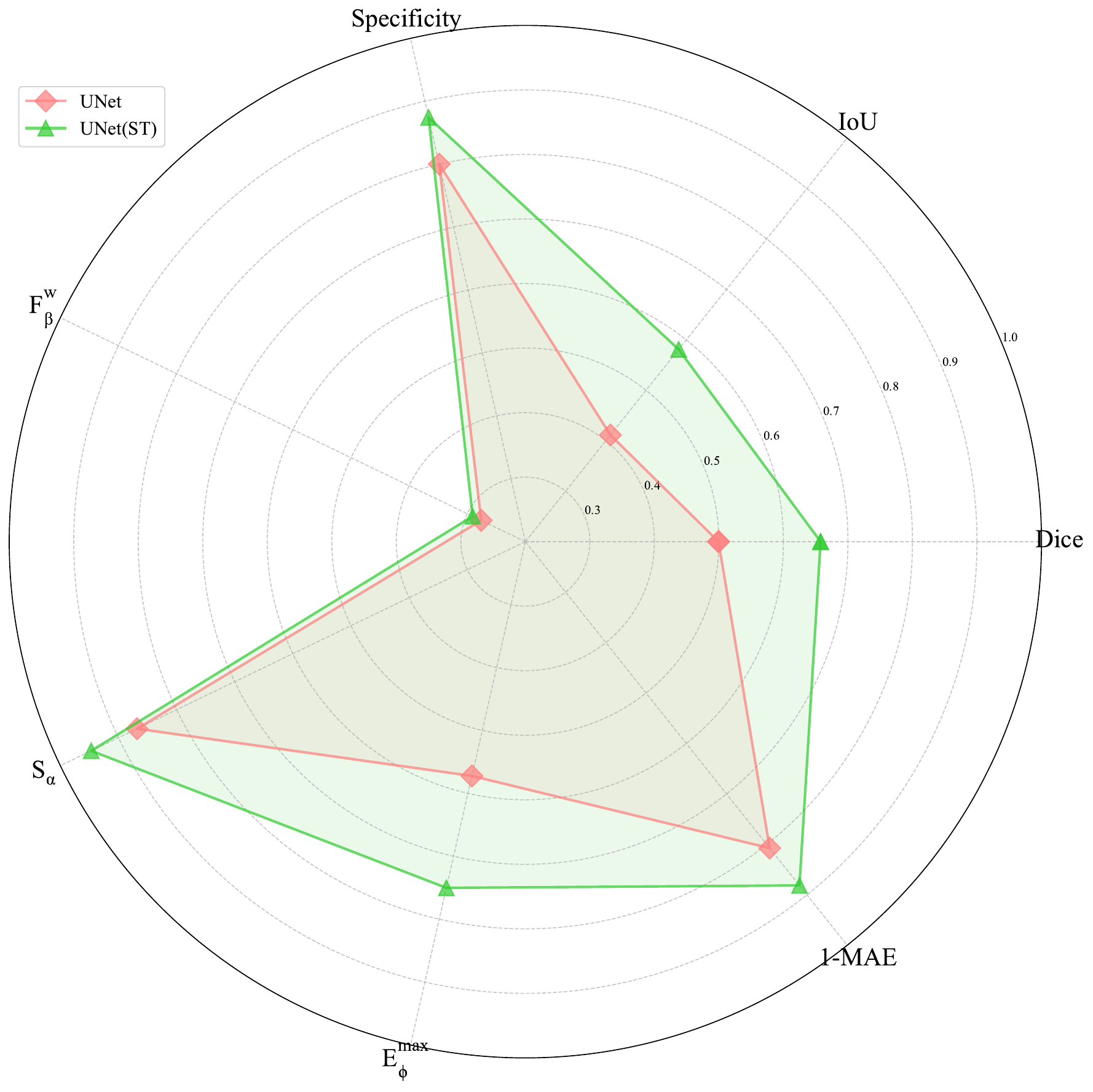}
        \label{fig:figure_Radar4}
    \end{minipage}\hfill
    \begin{minipage}{0.28\linewidth}
        \centering
        \includegraphics[width=\linewidth]{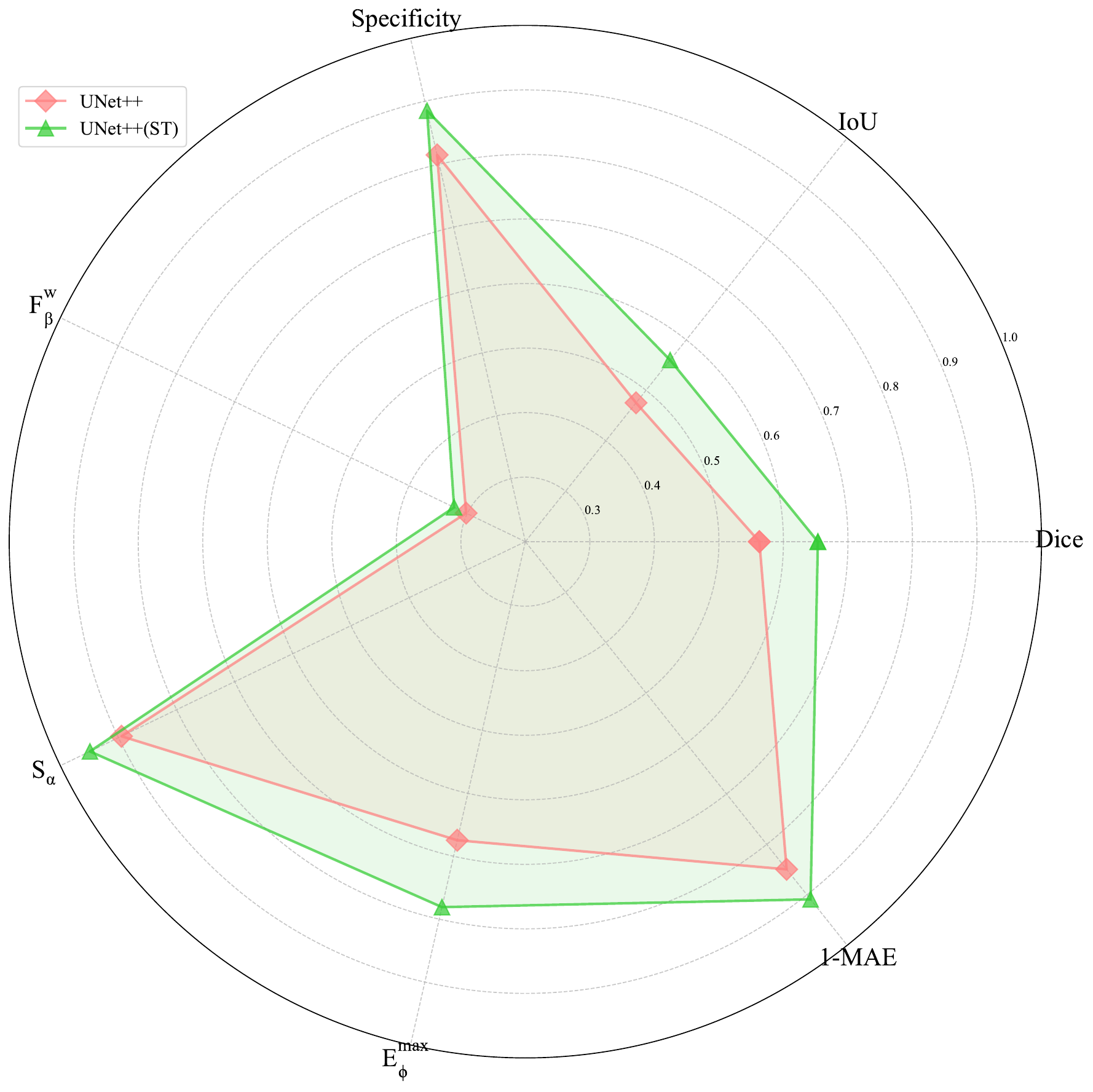}
        \label{fig:figure_Radar5}
    \end{minipage}\hfill
    \begin{minipage}{0.28\linewidth}
        \centering
        \includegraphics[width=\linewidth]{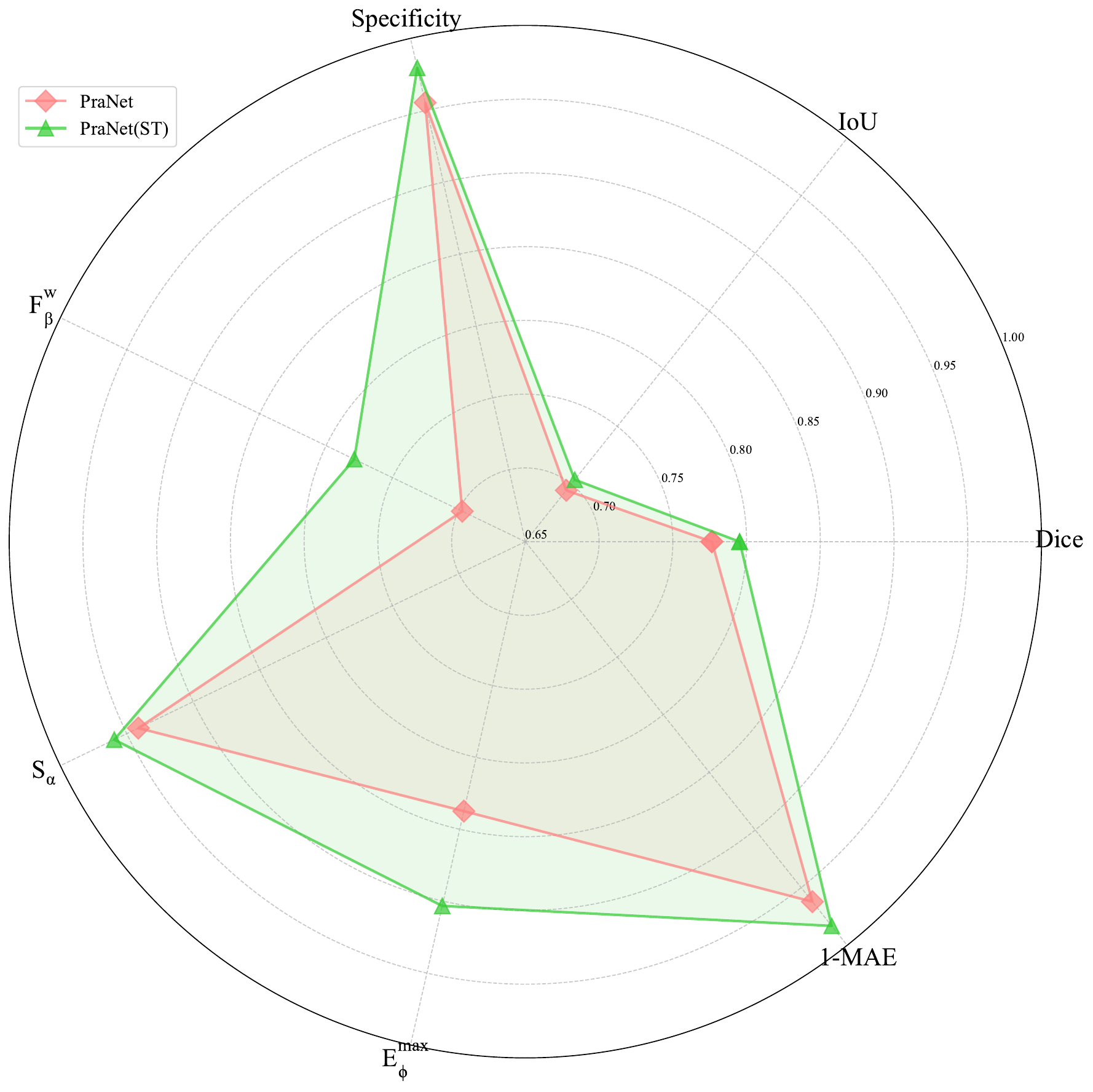}
        \label{fig:figure_Radar6}
    \end{minipage}
    \caption{Radar charts illustrating the performance metrics (Dice, IoU, Specificity, $F_\beta^w$, $S_\alpha$, $E_\phi^{max}$, and 1-MAE) of UNet, UNet++ and PraNet segmentation models and their style transfer variants on Skin datasets.}
    \label{fig:figure_Radar_Skin}
\end{figure*}

From Table \ref{tab:performance2}, we can observe that performing style transfer prior to segmentation improves various metrics on the test set compared to segmentation models trained on the original images. Specifically, across different image segmentation models, the Dice coefficient, IoU (Intersection over Union), and $F_\beta^w$ increased by approximately 10\%. Moreover, as illustrated in Figure \ref{fig: results2}, qualitative analysis demonstrates that applying style transfer to the test set results in better-performing segmentation models, with reduced noise and a lower false positive rate. Furthermore, we have additionally created radar charts \ref{fig:figure_Radar_Skin} to more effectively highlight the performance disparities among the various models.

\begin{table}[htbp]
\centering
\adjustbox{width=\linewidth}{%
\begin{tabular}{|c|c|c|c|c|c|c|c|}
\toprule
Model & Dice & IoU & Specificity & $F_\beta^w$ & $S_\alpha$ & $E_\phi^{max}$ & MAE\\
\midrule
UNet & 0.4994 & 0.4114 & 0.8001 & 0.2760 & 0.8680 & 0.5724 & 0.1926\\
UNet(\textbf{ST}) & \textbf{0.6576} & \textbf{0.5811} & \textbf{0.8743} & \textbf{0.2907} & \textbf{0.9469} & \textbf{0.7503} & \textbf{0.1186} \\
\cmidrule{1-8}
UNet++ & 0.5627 & 0.4752 & 0.8148 & 0.3019 & 0.8949 & 0.6746 & 0.1505 \\
UNet++(\textbf{ST}) & \textbf{0.6534} & \textbf{0.5599} & \textbf{0.8850} & \textbf{0.3226} & \textbf{0.9491} & \textbf{0.7809} & \textbf{0.0910}\\
\cmidrule{1-8}
PraNet & 0.7765 & 0.6945 & 0.9553 & 0.6975 & 0.9413 & 0.8374 & 0.0378 \\
PraNet(\textbf{ST}) & \textbf{0.7953} & \textbf{0.7037} & \textbf{0.9795} & \textbf{0.7788} & \textbf{0.9594} & \textbf{0.9033} & \textbf{0.0168} \\
\bottomrule
\end{tabular}}
\caption{Comparison results of different models and their style-transferred versions in terms of performance indicators.}
\label{tab:performance2}
\end{table}

\section{Conclusion}
\label{sec:conclusion}
Our proposed method effectively addresses the challenges posed by domain shifts in medical image segmentation. By transforming images from various domains into a unified style while preserving critical anatomical structures, we enhance the robustness and accuracy of segmentation models. The approach does not require the target domain to be included in the training set, making it practical for real-world applications where acquiring diverse training data is challenging. Additionally, the method is versatile and can be integrated with existing segmentation models, providing a valuable tool for improving performance in diverse clinical settings. Future efforts will focus on how to directly devise a loss metric that evaluates the quality of image segmentation after style transfer, as well as on enabling models to achieve accurate medical image segmentation directly when trained on stylized datasets.

\end{document}